\documentclass[10pt]{article}
\usepackage{subfig}
\usepackage{titlesec}
\usepackage{esdiff}
\usepackage{amsmath}
\usepackage{amssymb}
\usepackage{authblk}
\usepackage{bbm}
\usepackage{float}

\usepackage[normalem]{ulem}
\usepackage{bbold}
\usepackage[colorlinks=true,linkcolor=blue,citecolor=teal,urlcolor=teal]{hyperref}%
\usepackage{geometry}
\usepackage{BOONDOX-cal}
\usepackage{setspace}
\usepackage{cite}
\usepackage{flushend}
\usepackage{mathrsfs}
\usepackage{hyperref}
\usepackage{graphicx}
\usepackage{cancel}
\usepackage{braket}
\usepackage{array,esint}
\usepackage[most]{tcolorbox}

\def\s{\sigma}

\def\t{\tau}

\newcommand{\be}{\begin{equation}}
\newcommand{\ee}{\end{equation}}
\newcommand{\bes}{\begin{equation*}}
\newcommand{\ees}{\end{equation*}}

\NewEnviron{eqn}{
\begin{align}
\begin{split}
  \BODY
\end{split}
\end{align}
}

\NewEnviron{eqn*}{
\begin{align*}
\begin{split}
  \BODY
\end{split}
\end{align*}
}

\graphicspath{}
\titleformat{\section}{\fontsize{12}{12}\bfseries}{\thesection}{1em}{}
\twocolumn
\begin{document}
\twocolumn[\begin{@twocolumnfalse}
\title{\textbf{Virtual transitions in an atom-mirror system
in the presence of two scalar photons}}
\author{\textbf{Ashmita Das${}^{a*}$, Soham Sen${}^{a\dagger}$ and Sunandan Gangopadhyay${}^{a\ddagger}$}}
\affil{{${}^a$ Department of Astrophysics and High Energy Physics,}\\
{S.N. Bose National Centre for Basic Sciences,}\\
{JD Block, Sector III, Salt Lake, Kolkata 700106, India}}
\date{}
\maketitle
\begin{abstract}
\noindent 
We examine the virtual transition of an atom-mirror system with the simultaneous emission of two scalar photons, where the atom and the mirror admit a  relative acceleration between them. For the single photon emission, literature (\href{https://link.aps.org/doi/10.1103/PhysRevLett.121.071301}{Phys. Rev. Lett. 121 (2018) 071301}) dictates that  the transition probabilities of two individual systems, such as an atom accelerating with respect to the mirror and its reverse, turn out to be equivalent under the exchange of the frequencies of atom and the field. Addressing the observational merit of such excitation process, a detectable probability ($P \sim 10^{-2}$) is also reported in the above literature. 
 In the present manuscript our finding dictates that the 
simultaneous emission of dual photon instead of one, destroys the equivalence between the transition probabilities as reported in the above literature. 

\end{abstract}
\end{@twocolumnfalse}]
\section{Introduction}\label{intro}
\noindent\let\thefootnote\relax\footnote{{}\\
{$*$ashmita.phy@gmail.com}\\
{$\dagger$sensohomhary@gmail.com, soham.sen@bose.res.in}\\
{$\ddagger$sunandan.gangopadhyay@gmail.com}} For the past several years we have seen a rich amalgamation of several individual disciplines of physics, which led us to many interesting phenomena. 
The classic example is none other than the combination of  general relativity\cite{Einstein15,Einstein16} and thermodynamics in the context of black holes and cosmological spacetime \cite{Hawking,Hawking2,Hawking3,Bekenstein,Bekenstein2,Page,Page2,Page3,Fulling21,Davies,DeWitt}.
Furthermore in recent times 
the blending of quantum optics and the acceleration radiation in flat/curved spacetime became somewhat popular and considered to be an alternative mechanism in explaining the Unruh-Fulling (UF) effect  \cite{Scully2003zz, Fulling2,Fulling,Ordonez1,Ordonez2,Ordonez3,Ordonez4,OTM, OTM2}. 
In the context of black hole spacetime this approach predicts the presence of an entropy which has different origin than that of the Bekenstein-Hawking entropy, named as horizon brightened acceleration radiation (HBAR) entropy \cite{Fulling2}. 
More along this line is the use of cavity quantum electrodynamics in the study of acceleration radiation which shows that the transition probability of an accelerated atom can be increased to many orders in magnitude than we usually obtain in the standard Unruh radiation \cite{Scully2003zz}. 
%
%
%
Since the proposition of UF effect, it has proven its prominence in case of the Hawking radiation in black hole spacetime \cite{Hawking,Hawking2}, particle emission by cosmological horizon \cite{GWGibbons}, Schwinger mechanism \cite{Schwinger}, quantum entanglement \cite{Nielsen}, etc. 
The idea of UF effect dictates that in Minkowski spacetime a uniformly accelerating observer perceives the vacuum state of a field as a thermal bath in equilibrium with temperature $T=\frac{\hbar a}{2\pi k_B c}$. Here $k_B,\,c,\, \hbar$ denote the Boltzmann constant, speed of light in vacuum and the reduced Planck's constant respectively, and $a$ symbolises the acceleration of the observer \cite{Unruh21, Unruh22}.  The observational aspects of UF effect are widely explored by modelling a uniformly accelerated two level atomic detector, known as  Unruh-DeWitt detector (UD) \cite{Unruh2,Higuichi,Cosmospace}. 
The construction of UD detector states that, in flat spacetime, a point like uniformly accelerated detector records particle excitation due to the interaction with a scalar field, located in its Minkowski vacuum. It captures the main essence of the UF effect in terms of the transition rate/ power spectrum of the UD detector. For some brief review on UF effect and modelling of UD detector, we refer our readers to \cite{Birrell,Higuichi}.
At this stage we resume the discussion related to the alternative approaches in the studies of acceleration radiation/UF effect.  It is well known that in quantum theory of fields, the virtual processes due to the vacuum fluctuations have disclosed many new quantum phenomena such as Lamb shift in the hydrogen atom which led to the construction of quantum electrodynamics, Raman scattering in the field of spectroscopy etc.
 In terms of the atom-field interaction it dictates that a two level atom makes a transition to its excited state while a virtual photon is simultaneously emitted. Subsequently the atom promptly comes back to its ground state by absorbing the photon.   
It was proposed in \cite{Scully2003zz} that the UF effect can also be perceived by interrupting a virtual process, where the emitted virtual photon turns into the real observable photon. 
Due to the accelerated motion of the atom, it get away from its original point of virtual emission which yields a nonzero probability for the atom not absorbing the emitted virtual photon. This transforms a virtual photon into a real one in the final state of the system \cite{Scully2003zz,Fulling2}.
 It is worth to be mentioned that the atom acquires acceleration by extracting energy from some external force agency which drives the centre of mass motion of the atom \cite{Scully2003zz,Fulling2,Fulling}.  This explains the notion of accelerated radiation/UF effect via this alternative approach. 
This mechanism has unveiled many interesting findings which can be found in the following references \cite{Fulling2, Fulling, OTM,OTM2}.
Using this alternative mechanism, the authors of \cite{Fulling}, have considered a system which  contains an atom and a mirror with a relative acceleration between them. They have examined two possible setup such as the atom is uniformly accelerating with respect to a fixed mirror and the reverse. Subsequently, the transition probabilities for both the systems are obtained due to the virtual processes.
It was shown that the transition probabilities are different from each other, however,  
both the probabilities are related under the exchange of atom and field frequencies.
 The authors have defined this connection as the ``application of equivalence principle in QED". 
 Moreover a spatial oscillatory behaviour of the transition probabilities has been reported in \cite{Fulling} 
 due to the interference of the ``incident" and ``reflected" photons. Note that this oscillatory feature cannot be removed by integrating out the probabilities over all values of field frequency.  
 In relation to the above equivalence, we mention that a violation of this equivalence relation has recently been reported in the background of a generalized uncertainty principle framework \cite{SG}.
For experimental verification of such excitations, the authors of \cite{Fulling} have proposed a setup where an ensemble of $N$ polar molecules is coupled with a superconducting transmission line
microwave cavity terminated by a SQUID (semiconducting quantum interference device) \cite{wilson_nature,Flat1,Flat2}. This setup mimics an accelerating mirror and two level atomic system \cite{Fulling}. It was shown  that for all $N \sim 10^{4}-10^{6}$ polar molecules and the acceleration $(a)$ of the mirror larger than the $c$ times the field frequency ($c \nu$), the transition probability lies within the detectable range ({\it i.e} $P_{exc} \sim 10^{-2}$).  
The observational merit of this model gives us a strong motivation to extend the model to the simultaneous emission of two scalar photons instead one. 

\noindent
In recent times the authors of \cite{Das} have proposed a toy model of UD detector simultaneously interacting with multiple scalar fields. This work has  revealed some interesting features in terms of the transition probability of the detector which cannot be achieved in case of an atom and single field interaction. 
Following \cite{Das}, we consider the interaction Lagrangian as $\sim \s(\t)\phi_1[x(\t)]\, \phi_2[x(\t)]$ for the simultaneous emission of two photons. Here $\phi_1[x(\t)],\phi_2[x(\t)]$ depict two scalar photon fields and $\s(\t)$ represents the operators in detector sector. $\t$ denotes the proper time of the atomic detector. 
At this stage we briefly mention the origin of such interaction Lagrangian \cite{Das}.   The standard interaction term of a UD detector and single scalar field is of the form, $\mathcal{L}_{int}\sim \s({\t}) \phi[x(\t)]$. 
When the field makes a transition from its vacuum to 1-particle state
and the detector moves to its energy eigenstate $\ket{E}_f$ from $\ket{E}_i$,  a nonzero transition amplitude can be achieved within the first order in perturbation theory.  The transition amplitude for such system becomes as follows \cite{Birrell, Higuichi}, 
\be
  A(E_f|E_i) \sim \bra{E_f} \s(\t) \ket{E_i} \bra{1_p} \phi[x(\t)] \ket{0}~,
 \label{standard amplitude}
 \ee
Following the single field interaction one can intuitively write the interaction Lagrangian for the dual photons emission as, 
\be
 \mathfrak{L}_{int} =\mathbbm{g}\, \s(\t)(\phi_1[x(\t)]+ \phi_2[x(t)])~,
 \label{scalar-2}
\ee  
 Note unlike single field case, now we have several choices of final states due to the emission of two fields. Below we briefly discuss such two possible cases. For detailed discussion we refer our readers to the section.(VI) of \cite{Das}

\noindent
Case 1: One of the possible final state can be written as, 
 \begin{eqn}
\ket{\small{\mbox{initial}}} &= \ket{E_i} \otimes \ket{0_1} \otimes \ket{0_2} ~;\\
\ket{\small{\mbox{final}}} &= \ket{E_f} \otimes \Bigl( \ket{1_{p_1}} \otimes \ket{0_2}  +  \ket{0_1} \otimes \ket{1_{p_2}}  \Bigr)~.
\label{states-1}
\end{eqn}
Here $\ket{0_{1,2}}$ and  $\ket{1_{p_{1,2}}}$ depict the vacuum and 1-particle state for each field. In this scenario the detector is interacting with two scalar fields where at a particular instant a single field interacts with the detector and creates 1-particle state, while another is lying in its vacuum.
Therefore in this case the transition amplitude varies as the summation of the two transitions {\it i.e} $\sim (A_1 + A_2)$, where $A_1,\, A_2$ denote the amplitudes as similar to eq.(\ref{standard amplitude}). 
This case has its own merits and may reveal interesting features \cite{Das}.

\noindent
Case 2:  In this case both the fields make simultaneous transition to their respective 1-particle states due to the interaction with the detector at the same point in spacetime.  This leads to the following final state of the system :  
\be
 \ket{\small{\mbox{final}}} = \ket{E_f} \otimes \ket{1_{p_1}} \otimes \ket{1_{p_2}} ~,
\label{final_2}
 \ee
when the initial state is same as written in Eq.(\ref{states-1}). 
For the interaction Lagrangian in Eq.(\ref{scalar-2}), initial and final states as in Eqs.(\ref{states-1}) and (\ref{final_2}) respectively, a nonzero transition amplitude can be achieved, if one considers the higher order term in perturbation theory  {\it i.e}, $\sim \bra{\small{\mbox{final}}} \mathfrak{L}_{int}^2 \ket{\small{\mbox{initial}}}$. 
The term $\mathfrak{L}_{int}^2$, will bring in the following combination of fields, 
 \be
 \sim \phi_1^2 + \phi_2^2 + 2 \phi_1 \phi_2~.
 \label{cross_1}
 \ee
Note that the interaction like $\sim \phi_1^2$ and $\sim \phi_{2}^{2}$ contribute to the transition amplitude when the fields make a transition from the vacuum to their 2-particle state. In the present work we restrict our analysis within the transition of the fields to 1-particle states as energetically 1-particle state is most favoured \cite{Das}. 
Therefore in the present context, only the third term of Eq.(\ref{cross_1}) contribute to the transition amplitude. This analysis implies that in working with the Lagrangian $\sim \phi_1[x(\t)] \phi_2 [x(\t)]$ is not an arbitrary choice, rather it originates from a fundamental interaction such as Eq.(\ref{scalar-2}). 
The emergence of various final states
and interaction Lagrangian are solely due to the dual field interaction and cannot be achieved in single field setup. Therefore these features render the distinctive characters of single and dual field emission processes.   

\noindent
With respect to the previous literature, it is seen that apart from the field theoretic interests,  multiple fields analyses have earned loads of attentions in the sector of beyond the standard model of particle physics, cosmology, dark matter physics, etc.  
 It is well known that in search for the signatures of extra dimensions in the collider experiments (such that Large Hadron Collider), the Kaluza-Klein (KK) modes of bulk graviton and bulk standard model are the potential candidates \cite{Davoudiasl:1999jd, Chang:1999nh}. 
 These KK modes are originated due to the compactification of the spatial extra dimensions and interpreted as the multiple fields in 4-dimensional effective theory \cite{Dienes:2002hg}. 
 %
 %
Furthermore,  in \cite{Dvali:2007hz},  a large number of elementary particle species have been considered in resolving the hierarchy problem of the two fundamental length scales such as  electroweak and Planck scale. 
 Later, it was shown that an appropriate abundance of dark matter can also be achieved while considering a large number of copies of the standard model in \cite{Dvali:2009ne, Dvali:2009fw}.
In cosmology, the multiple field inflationary models are considered to explore the inflationary dynamics, inflationary trajectories and particle production \cite{Wands:2007bd, Gong:2016qmq, Alvarez-Gaume:2011wmd}. Two-field inflationary scenario have been studied in the context of swampland de Sitter conjecture \cite{NooriGashti:2021nox}, reheating in cosmology \cite{Martin:2021frd}, dark energy, \cite{Akrami:2020zfz} etc. These promising features of multiple field theories instigate us to explore the implications of multi-field interaction with two level atomic detector in the present manuscript.
 Thus we propose a system which contains a two-level atom and a mirror with an existent relative acceleration between them. The system becomes excited due to the virtual processes in the vacuum states of the photon fields. However, due to the relative acceleration of the atom and  mirror the two real scalar photons are simultaneously emitted and detected at the same time. Concurrently, The atom makes a transition to its higher excited state.
 We make an attempt to map our system within the experimental setup as described in \cite{Fulling} as following.  Within a cavity setup an atom makes a transition to its higher excited state while simultaneously two photons are emitted with two different single frequency modes due to the interaction of the atom and dual photon. Thus in contrast to the single field emission process, one obtains two quanta of detectable photons carrying two different frequencies within a cavity. 
 We examine the transition probabilities of such systems under the conditions where the atom(mirror) is accelerating with respect to fixed mirror (atom). 
We enlist our findings as below, 
\begin{itemize}
\item
 The so called ``equivalence principle in QED"\cite{SG,Philbin,Rohrlich,Vallisneri,Singleton,Singleton2,Fulling0} is no more holding in case of the dual field emission process. 
 This solely happens due to the emergence of a cross term in the transition amplitude for the emission of dual photon modes. 
 \item
 We achieve a correction term as a function of $(a, \,\nu,\, \omega)$, over the single field emission case, where, $\omega$ represents the angular frequency of the atom. This correction term plays an important role as we observer that for an accelerating mirror (fixed atom) system the transition probability becomes a nontrivial function of $(a,\,\nu)$ even under the condition $a \gg c \nu$. This is in contrast with the outcome in \cite{Fulling}. 
 %
 %
 %
 %
 Undoubtedly, this outcome promotes more investigations concerning with the atomic detector and multiple fields interaction and also signifies the observational prominence of the present setup. 
 \end{itemize}
We organise our manuscript as follows. In section(\ref{moving_atom}, \ref{moving_mirror}), we briefly describe the two systems and obtain the corresponding transition probabilities. In section(\ref{equivalence}), we study the relation between the excitation probabilities of these two systems and subsequently discuss the phenomenological aspects of our model in section(\ref{pheno}) We discuss our result in section(\ref{discussion}).  

\section{A uniformly accelerating atom and a fixed mirror}
\label{moving_atom}
We consider a system which contains a uniformly accelerated two level atom with respect to a fixed mirror.  
The atom follows a trajectory as below, 
\begin{align}
t(\tau)=&\frac{c}{a}\sinh\left(\frac{a\tau}{c}\right)\,\,\,\,\,\,\,\,\,\,\,\,\,\,\,\,
z(\tau)=&\frac{c^2}{a}\cosh\left(\frac{a\tau}{c}\right)\label{1.3},
\end{align}
where $(a, \,\tau)$ symbolise the uniform acceleration of the atom along the $z$-direction and the proper time of the atom respectively. 
The atom is interacting with two scalar photons at the same time and possesses a transition angular frequency $\omega$.  
 We restrict the wave vectors of the two photons ($\mathbf{k}_1,\mathbf{k}_2$) to be parallel to the $z$-axis and the frequencies of the same are denote by $\nu_1$ and $\nu_2$.
In the initial state of the system the atom lies in its ground state while the fields are located in their respective Minkowski vacuum. Thus the initial state can be written as : 
 $\ket{\mbox{initial}}= \ket{g} \otimes \ket{0_1} \otimes \ket{0_2}$, 
 where $\ket{g}$ is the ground state of the atom and the $\ket{0_{1,2}}$ stand for the vacuum states of two scalar photons.  Following the discussion in \cite{Das}, it can be perceived that the simultaneous emission of two photons leads to a final state of the system as below, 
  \begin{equation}
 \ket{\small{\mbox{final}}} = \ket{e} \otimes \ket{1_{\nu_1}} \otimes \ket{1_{\nu_2}} ~.
  \end{equation}
 Here $
 \ket{e}$ represents the excited state of the atom.  Allowing upto the first order in perturbation theory, a nonzero transition amplitude can be obtained only for the choice of interaction Hamiltonian as following, 
\begin{equation}\label{1.5}
\hat{\mathcal{H}}_I(\tau)=\hbar \mathcal{G}\left[\hat{b}_{\nu_1}\phi_{\nu_1}+h.c.\right]\left[\hat{b}_{\nu_2}\phi_{\nu_2}+h.c.\right]\left[\hat{\sigma}e^{-i\omega\tau}+h.c.\right]
\end{equation}
Here $\mathcal{G}$ denotes the atom-fields coupling constant, $\phi_{\nu_{1,2}}(x(\tau))$ are the mode solutions of two scalar photon fields and ($\hat{\sigma},\,\hat{\sigma}^{\dagger}$) stand for the lowering and raising operators of the atom. $\hat{b}_{\nu_{1,2}}$ 
denote the annihilation operators corresponding to the two scalar photons. 
We assume that the mirror is fixed at $z_0(<\frac{c^2}{a})$ along the $z$-axis which yields the following boundary condition for the mode solutions, 
\begin{equation}\label{1.6}
\phi_{\nu_{1,2}}\biggr|_{z=z_0}=0~.
\end{equation}
Eq.(\ref{1.6}) dictates that the normal modes of the photon fields ought to be the superposition of the incident and the reflected waves which in turn produces standing waves as below,  
\begin{equation}\label{1.7}
\phi_{\nu_{1,2}}=e^{-i\nu_{1,2}t-ik_{1,2}(z-z_0)}-e^{-i\nu_{1,2}t+ik_{1,2}(z-z_0)}~.
\end{equation}  
We take the duration of atom-fields interaction to be infinite and thus the transition amplitude for this excitation can be written as, 
\begin{equation}\label{1.11}
\begin{split}
\mathcal{A}_{|i\rangle\rightarrow|f\rangle}=&-\frac{i}{\hbar}\int_{-\infty}^{+\infty}d\tau \langle 1_{\nu_1},1_{\nu_2},e|\hat{\mathcal{H}}_I(\tau)|0_{\nu_1},0_{\nu_2},g\rangle\\
=&-i\mathcal{G}\int_{-\infty}^{+\infty}d\tau~ e^{i(\nu_1+\nu_2)t}e^{i\omega\tau}\left(e^{ik_1(z-z_0)}-c.c.\right)\\
&\times\left(e^{ik_2(z-z_0)}-c.c.\right)
\end{split}
\end{equation} 
We replace $k_{1,2}=\frac{\nu_{1,2}}{c}$ in the above equation and
consider that the emitted scalar photons have identical frequencies {\it i,e } $\nu_1=\nu_2=\nu \implies\,k_1=k_2=\frac{\nu}{c}$.
Therefore the transition amplitude in eq.(\ref{1.11}) takes the following form,
\begin{equation}\label{1.12}
\mathcal{A}_{|i\rangle\rightarrow |f\rangle}=-i\mathcal{G}\int_{-\infty}^{\infty}d\tau~e^{2i\nu t} e^{i\omega\tau}\left(e^{\frac{2i\nu}{c}(z-z_0)}+c.c.-2\right)~.
\end{equation}
This yields the transition probability of the system as below,
\begin{equation}\label{1.13}
\begin{split}
\mathcal{P}_{exc}=\mathcal{G}^2\biggr|\int d\tau~e^{2i\nu t(\tau)} e^{i\omega\tau}\left(e^{\frac{2i\nu}{c}(z(\tau)-z_0)}+c.c.-2\right)\biggr|^2~.
\end{split}
\end{equation}
At this stage we use the trajectory for the accelerated atom as in Eq.(\ref{1.3}) in eq.(\ref{1.13}) and obtain the excitation probability as follows,
\begin{equation}\label{1.14}
\begin{split}
\mathcal{P}_{exc}=&\mathcal{G}^2\biggr|\int_{-\infty}^{+\infty}d\tau \biggr(e^{\frac{2i\nu c}{a}e^{\frac{a\tau}{c}}-2ikz_0}+e^{-\frac{2i\nu c}{a}e^{-\frac{a\tau}{c}}+2ikz_0}\\&-2e^{\frac{2i\nu c}{a}\sinh\left(\frac{a\tau}{c}\right)}\biggr)e^{i\omega\tau}\biggr|^2
\end{split}
\end{equation}
Now we evaluate the three integrals in Eq.(\ref{1.14}) separately where the first integral is given by,
\begin{equation}\label{1.15}
I_1=\int_{-\infty}^{+\infty}d\tau~e^{\frac{2i\nu c}{a}e^{\frac{a\tau}{c}}-2ikz_0}e^{i\omega\tau}~.
\end{equation}
Changing the variable such that $x=\frac{2\nu c}{a}e^{\frac{a\tau}{c}}$, Eq.(\ref{1.15}) reads,
\begin{equation}\label{1.17}
\begin{split}
I_1=&\frac{c}{a}\left(\frac{a}{2\nu c}\right)^{\frac{i\omega c}{a}}\int_{0}^{\infty}dx ~e^{ix}x^{\frac{i\omega c}{a}-1}e^{-2ikz_0}\\
=&\frac{c}{a}\left(\frac{a}{2\nu c}\right)^{\frac{i\omega c}{a}}e^{-2ikz_0}e^{-\frac{\pi c \omega}{2a}}\Gamma\left[\frac{i\omega c}{a}\right]~.
\end{split}
\end{equation}
Following the similar change of variables the 
second integral in eq.(\ref{1.14}) yields, 
%
%
\begin{equation}\label{1.19}
\begin{split}
I_2=&\frac{c}{a}\left(\frac{a}{2\nu c}\right)^{-\frac{i\omega c}{a}}\int_{0}^{\infty}dx' ~e^{-ix'}{x'}^{-\frac{i\omega c}{a}-1}e^{2ikz_0}\\
=&\frac{c}{a}\left(\frac{a}{2\nu c}\right)^{-\frac{i\omega c}{a}}e^{2ikz_0}e^{-\frac{\pi c \omega}{2a}}\Gamma\left[-\frac{i\omega c}{a}\right]~.
\end{split}
\end{equation}
The third integral, which solely appears due to the dual field interaction in eq.(\ref{1.14}) is given by, 
\begin{equation}\label{1.20}
I_3=-2\int_{-\infty}^{+\infty}d\tau~e^{\frac{2i\nu c}{a}\sinh\left(\frac{a\tau}{c}\right)}e^{i\omega\tau}~.
\end{equation}
Under the coordinate transformation $\tilde{x}=e^{\frac{a\tau}{c}}$, $I_3$ becomes, 
\begin{equation}\label{1.21}
\begin{split}
I_3=&-\frac{2c}{a}\int_0^\infty d\tilde{x}~ {\tilde{x}}^{\frac{i\omega c}{a}-1}e^{\frac{i\nu c}{a}\left(\tilde{x}-\frac{1}{\tilde{x}}\right)}~.
\end{split}
\end{equation}
Performing the above integration 
the final form of $I_3$ is obtained as follows \cite{Gradshteyn,HAMLD}, 
\begin{equation}\label{1.22}
I_3=-\frac{4c}{a}e^{-\frac{\pi\omega c}{2a}}K_{\frac{i\omega c}{a}}\left(\frac{2\nu c}{a}\right),
\end{equation}
where $K_{\frac{i\omega c}{a}}\left(\frac{2\nu c}{a}\right)$ is the modified Bessel function. 
Combining the Eq.(\ref{1.17}) and Eq.(\ref{1.19}), we obtain
\begin{equation}
I_1+I_2=\frac{c}{a}e^{-\frac{\pi \omega c}{2a}}\left(\left(\frac{a}{2\nu c}\right)^{\frac{i\omega c}{a}}e^{-2ikz_0}~\Gamma\left[\frac{i\omega c}{a}\right]+c.c.\right)~.
\label{1.23}
\end{equation}
For a complex parameter $z$ and its complex conjugate $z^*$, one obtains,  $z+z^*=2\Re(z)$.  One can also write the complex number $z$ as, $z=|z|\left[\cos(\text{arg(z)})+i\sin(\text{arg(z)})\right]$. Using these relations, we recast the sum in eq.(\ref{1.23}) as,
\begin{equation}\label{1.24}
\begin{split}
I_1+I_2=&\frac{2c}{a}e^{-\frac{\pi c\omega}{2a}}\Re\left(\left(\frac{a}{2\nu c}\right)^{\frac{i\omega c}{a}}e^{-2ikz_0}~\Gamma\left[\frac{i\omega c}{a}\right]\right)\\
=&\frac{2c}{a}e^{-\frac{\pi c\omega}{2a}}\Re\biggr[\biggr(\cos\left(\frac{2\nu z_0}{c}-\frac{\omega c}{a}\ln\left[\frac{a}{2\nu c}\right]\right)\\-&i\sin\left(\frac{2\nu z_0}{c}-\frac{\omega c}{a}\ln\left[\frac{a}{2\nu c}\right]\right)\biggr)\biggr(\Re\left[\Gamma\left[\frac{i\omega c}{a}\right]\right]\\+&i\Im\left[\Gamma\left[\frac{i\omega c}{a}\right]\right]\biggr)\biggr]\\
=&\frac{2ce^{-\frac{\pi c\omega}{2a}}}{a}\biggr[\cos\left[\frac{2\nu z_0}{c}-\frac{\omega c}{a}\ln\left[\frac{a}{2\nu c}\right]\right]\Re\left[\Gamma\left[\frac{i\omega c}{a}\right]\right]\\+&\sin\left[\frac{2\nu z_0}{c}-\frac{\omega c}{a}\ln\left[\frac{a}{2\nu c}\right]\right]\Im\left[\Gamma\left[\frac{i\omega c}{a}\right]\right]\biggr]~.
\end{split}
\end{equation}
In the above equation we substitute the following relations,
\begin{align}
&\Re\left[\Gamma\left(\frac{i\omega c}{a}\right)\right]=\left|\Gamma\left(\frac{i\omega c}{a}\right)\right|\cos\left(\text{arg}\left[\Gamma\left(\frac{i\omega c}{a}\right)\right]\right)\label{1.25}\\
\nonumber\\
&\Im\left[\Gamma\left(\frac{i\omega c}{a}\right)\right]=\left|\Gamma\left(\frac{i\omega c}{a}\right)\right|\sin\left(\text{arg}\left[\Gamma\left(\frac{i\omega c}{a}\right)\right]\right)\label{1.26}~.
\end{align}
and obtain the final form of ($I_1+I_2$)  as,
\begin{equation}\label{1.27}
\begin{split}
I_1+I_2=&\frac{2c}{a}e^{-\frac{\pi c\omega}{2a}}\left|\Gamma\left(\frac{i\omega c}{a}\right)\right|\cos\left(\theta\right)
\end{split}
\end{equation}
where,  
\begin{equation}\label{1.28}
\theta=\frac{2\nu z_0}{c}-\frac{\omega c}{a}\ln\left[\frac{a}{2\nu c}\right]-\text{arg}\left[\Gamma\left(\frac{i\omega c}{a}\right)\right]~.
\end{equation}
We obtain the transition probability while summing up the Eqs.(\ref{1.22}, \ref{1.27}) as below, 
\begin{equation}\label{1.29}
\begin{split}
\mathcal{P}_{exc}=&\frac{4\mathcal{G}^2c^2}{a^2}e^{-\frac{\pi\omega c}{a}}\left|\Gamma\left(\frac{i\omega c}{a}\right)\right|^2\cos^2(\theta)\biggr|\biggr(1\\-&2\sec(\theta)\left|\Gamma\left(\frac{i\omega c}{a}\right)\right|^{-1}K_{\frac{i\omega c}{a}}\left(\frac{2\nu c}{a}\right)\biggr)\biggr|^2\\
=&\frac{8\pi\mathcal{G}^2c}{a\omega}\frac{\cos^2(\theta)}{e^{\frac{2\pi\omega c}{a}}-1}\biggr(1-\frac{4\sec(\theta)}{\left|\Gamma\left(\frac{i\omega c}{a}\right)\right|}K_{\frac{i\omega c}{a}}\left(\frac{2\nu c}{a}\right)\\+&\frac{4\omega c \sec^2(\theta)}{a\pi}\biggr|K_{\frac{i\omega c}{a}}\left(\frac{2\nu c}{a}\right)\biggr|^2\sinh\left(\frac{\pi\omega c}{a}\right)\biggr)
\end{split}
\end{equation}
Here, $K_{\frac{i\omega c}{a}}\left(\frac{2\nu c}{a}\right)$ is a real number provided $\nu,c,a$ and $\omega$ are real. 
Note that the excitation probability in eq.(\ref{1.29}) depends on the parameters, $\mathcal{G},a,\nu,\omega$ and $z_0$. 
We explore the relevant phenomenology of this result in the section (\ref{pheno}). 
\section{A uniformly accelerating mirror and a fixed atom}
\label{moving_mirror}
A uniformly accelerating mirror with respect to a fixed atom is considered in this section. 
As similar to the section(\ref{moving_atom}), photons are simultaneously produced from their respective vacua due to the accelerated motion of the mirror and subsequently detected by the atom at the same time in Minkowski spacetime. With the simultaneous detection of the two field quanta, the atom gets excited and jumps to its higher energy state. 
The atom is fixed at $z=z_0<\frac{c^2}{a}$ in Minkowski spacetime.
At this stage we take a widely used coordinate transformation such as,  
\begin{align}
t=&\frac{c}{a}e^{\frac{a\tilde{z}}{c^2}}\sinh\left(\frac{a\tilde{t}}{c}\right),\,\,\,\,\,\,\,\,\,\,\,
z=&\frac{c^2}{a}e^{\frac{a\tilde{z}}{c^2}}\cosh\left(\frac{a\tilde{t}}{c}\right)\label{1.31}~.
\end{align}
Here $(\tilde{t},\,\tilde{z})$ denote the spacetime coordinates for the moving mirror, ($t,\,z$) are the Minkowski coordinates and $a$ is a constant parameter.  When the mirror is following the trajectory such as 
$\tilde{z}=0$, it represents a uniformly accelerated mirror in the Minkowski spacetime with an acceleration $a$. 
This picture can be mapped to the Rindler space (defined in terms of coordinates $(\tilde {t}, \tilde{z})$), where one can realize this incident such as the mirror is at rest and atom is accelerating. Consequently, the vacuum state of the fields become Rindler vacuum which one defines with respect to the static mirror in Rindler space. On the other hand now the accelerated atom in Rindler space perceives particle production from the Rindler vaccum of the fields. Note that in Rindler space the Rindler vacuum does not perceive any horizon as it is defined with respect to the static mirror. Whereas the accelerated atom in Rindler space perceives  Rindler like horizon. 
Using Eq.(\ref{1.31}), the Minkowski line element ($ds^2=c^2dt^2-dz^2$) can be recasted in the form of the Rindler metric, which reads, 
\begin{equation}\label{1.32}
ds^2=e^{\frac{2a\tilde{z}}{c^2}}(c^2d\tilde{t}^2-d\tilde{z}^2)~.
\end{equation}
As the Rindler metric in Eq.(\ref{1.32}) is conformally flat to the Minkowski line element,  the positive frequency mode solutions of the scalar photons remain same as that of the standard mode solutions in Minkowski spacetime. Therefore in Rindler space the normal modes of the scalar photons can be expressed as standing waves, given by the following relation,
\begin{equation}\label{1.33}
\phi_{\nu_{1,2}}(\tilde{t},\tilde{z})=e^{-i\nu_{1,2}\tilde{t}+ik_{1,2}\tilde{z}}-e^{-i\nu_{1,2}\tilde{t}-ik_{1,2}\tilde{z}}
\end{equation} 
where $\nu_1$ and $\nu_2$ are the angular frequencies of photons 1 and 2 in the Rindler space \cite{Fulling}. %
The normal modes in Eq.(\ref{1.33})  depict the positive frequency mode solutions with respect to the Rindler time ($\tilde{t}$), which signifies that the vacuum states corresponding to these modes belong to the Rindler vacuum in Rindler space. 
%
%
%
%
 From Eq.(\ref{1.31}), we write the coordinates ($\tilde{t},\tilde{z}$) in terms of the ($t, z$) as below, 
\begin{align}
\tilde{t}=&\frac{c}{2a}\ln\left[\frac{z+ct}{z-ct}\right],\,\,\,\,\,\,\,\,\,\,
\tilde{z}=&\frac{c^2}{2a}\ln\left[\frac{a^2}{c^4}(z^2-c^2t^2)\right]~.\label{1.35}
\end{align}
Using Eq.(\ref{1.35}) in Eq.(\ref{1.33}), we achieve, 
\begin{equation}\label{1.36}
\begin{split}
\phi_{\nu_{1,2}}(t,z)=&e^{+\frac{i\nu_{1,2}c}{a}\ln\left[\frac{a}{c^2}(z-ct)\right]}\Theta[z-ct]\\-&e^{-\frac{i\nu_{1,2}c}{a}\ln\left[\frac{a}{c^2}(z+ct)\right]}\Theta[z+ct]
\end{split}
\end{equation}
where $\Theta$ denotes the Heaviside theta function. 
In line with the treatment as carried out in \cite{Fulling}, we consider that the mode solutions as written in Eq.(\ref{1.36}), in terms of Minkowski coordinates, initially contain no particles. This implies that these modes are in the Rindler vacuum of the corresponding fields which are defined with respect to the static mirror in Rindler space. 
Such a Rindler vacuum can be prepared outside a massive star just before it starts to collapse \cite{Fulling}. 
We carry our subsequent analysis in Rindler space and obtain the transition probability while using the trajectory of the atom as written in Eq.(\ref{1.35}). However, this analysis could have been done with a more physically realistic initial state in which the field modes are assumed to be in the Minkowski vacuum before they reflect off the mirror. A way to do this would be to follow the formalism in \cite{Birrell, Sue} which we shall not carry out in this paper. 
The mode solutions in Eq.(\ref{1.36})  portray that the right moving field modes can access only the region such that $z >\,c |t|$, which is known as right Rindler wedge (RRW). On the other hand left moving modes can only access the region $-z >\,c |t|$, called the left Rindler wedge (LRW). 
Individually the mode solution in RRW and LRW is not complete in whole Minkowski spacetime.  
However their combination as appears in Eq.(\ref{1.36}) is complete in all Minkowski spacetime and can be analytically continued to the future ($t > |z|/c$) and past ($-t > |z|/c$) regions.

Now we write the excitation probability of the system due to the simultaneous emission of two photons and the transition of atom to its higher excited state as below, 
\begin{equation}\label{1.37}
\mathcal{P}'_{exc}=\mathcal{G}^2\biggr|\int_{-\infty}^{+\infty}dt\, e^{i\omega t}\phi_{\nu_1}^*(t,z)\,\phi_{\nu_2}^*(t,z)\biggr|^2~.
\end{equation}
For simplicity we consider the two photons have equal angular frequency $\nu$. We substitute the position of the atom $z=z_0$ in the above equation, which yields,  
\begin{equation}\label{1.38}
\begin{split}
\mathcal{P}'_{exc}=&\mathcal{G}^2\biggr|\int_{-\infty}^{\infty}dt~ e^{i\omega t}\biggr(e^{-\frac{i\nu c}{a}\ln\left[\frac{a}{c^2}(z_0-ct)\right]}\Theta[z_0-ct]\\
&-e^{\frac{i\nu c}{a}\ln\left[\frac{a}{c^2}(z_0+ct)\right]}\Theta[z_0+ct]\biggr)^2\,\biggr|^2~.
\end{split}
\end{equation}
Note that $\Theta[z_0-ct]^2$ has non zero value only in the region $-\infty<t<\frac{z_0}{c}$ while $\Theta[z_0+ct]^2$ has non zero value in the region $-\frac{z_0}{c}<t<\infty$. For the cross-term,  $\Theta[z_0-ct]\Theta[z_0+ct]$, one obtains a non zero value in the regime $-\frac{z_0}{c}<t<\frac{z_0}{c}$. Using these conditions, we write the probability in eq.(\ref{1.38}) as, 
\begin{equation}\label{1.39}
\begin{split}
\mathcal{P}'_{exc}=&\mathcal{G}^2\biggr|\int_{-\infty}^\frac{z_0}{c}dt~e^{i\omega t}e^{-\frac{2i\nu c}{a}\ln\left[\frac{a}{c^2}(z_0-ct)\right]}\\
&+\int^{\infty}_{-\frac{z_0}{c}}dt~e^{i\omega t}e^{\frac{2i\nu c}{a}\ln\left[\frac{a}{c^2}(z_0+ct)\right]}\\
&-2\int_{-\frac{z_0}{c}}^\frac{z_0}{c}dt~e^{i\omega t}e^{\frac{i\nu c}{a}\ln\left[\frac{z_0+ct}{z_0-ct}\right]}\biggr|^2~.
\end{split}
\end{equation} 
For the first integral in the Eq.(\ref{1.39}), we change the variable $t\rightarrow-t$ and obtain, 
\begin{equation}\label{1.40}
\begin{split}
\mathcal{P}'_{exc}&=\mathcal{G}^2\biggr|\int_{-\frac{z_0}{c}}^{\infty}dt~e^{-i\omega t}e^{-\frac{2i\nu c}{a}\ln\left[\frac{a}{c^2}(z_0+ct)\right]}\\
&+\int^{\infty}_{-\frac{z_0}{c}}dt~e^{i\omega t}e^{\frac{2i\nu c}{a}\ln\left[\frac{a}{c^2}(z_0+ct)\right]}\\
&-2\int_{-\frac{z_0}{c}}^\frac{z_0}{c}dt~e^{i\omega t}e^{\frac{i\nu c}{a}\ln\left[\frac{z_0+ct}{z_0-ct}\right]}\biggr|^2~.
\end{split}
\end{equation}
Further we take $x=\omega\left(t+\frac{z_0}{c}\right)$ in the above equation, which leads us to the following form of $\mathcal{P}'_{exc}$. 
\begin{equation}\label{1.42}
\begin{split}
\mathcal{P}'_{exc}&=\mathcal{G}^2\biggr|\frac{1}{\omega}\int_{0}^{\infty}dx~e^{-i\left(x-\frac{\omega z_0}{c}\right)}x^{-\frac{2i\nu c}{a}}e^{-\frac{2i\nu c}{a}\ln\left[\frac{a}{c\omega}\right]}\\
&+\frac{1}{\omega}\int_{0}^{\infty}dx~e^{i\left(x-\frac{\omega z_0}{c}\right)}x^{\frac{2i\nu c}{a}}e^{\frac{2i\nu c}{a}\ln\left[\frac{a}{c\omega}\right]}\\
&-\frac{2e^{-\frac{i\omega z_0}{c}}}{\omega\left(\frac{2\omega z_0}{c}\right)^{\frac{i\nu c}{a}}}\int_0^{\frac{2\omega z_0}{c}}dx~e^{ix}x^{\frac{i\nu c}{a}}\left(1-\frac{cx}{2\omega z_0}\right)^{-\frac{i\nu c}{a}}\biggr|^2~.
\end{split}
\end{equation}
We denote the first two integrals in the above eq.(\ref{1.42}) as $I'_1$ and $I'_2$ and obtain the following relation
\begin{equation}\label{1.43}
\begin{split}
I'_1+I'_2&=-\frac{2\nu c}{a\omega}\left(\frac{a}{\omega c}\right)^{-\frac{2i\nu c}{c}}e^{\frac{i\omega z_0}{c}}e^{-\frac{\pi\nu c}{a}}\Gamma\left[-\frac{2i\nu c}{a}\right]\\
&-\frac{2\nu c}{a\omega}\left(\frac{a}{\omega c}\right)^{\frac{2i\nu c}{c}}e^{-\frac{i\omega z_0}{c}}e^{-\frac{\pi\nu c}{a}}\Gamma\left[\frac{2i\nu c}{a}\right]\\
&=-\frac{4\nu c}{a\omega}e^{-\frac{\pi\nu c}{a}}\left|\Gamma\left[-\frac{2i\nu c}{a}\right]\right|\cos\left(\theta'\right)
\end{split}
\end{equation}
where,
\begin{equation}\label{1.44}
\begin{split}
\theta'=&\frac{\omega z_0}{c}-\frac{2\nu c}{a}\ln\left[\frac{a}{\omega c}\right]+\text{arg}\left[\Gamma\left[-\frac{2i\nu c}{a}\right]\right]\\
=&\frac{\omega z_0}{c}-\frac{2\nu c}{a}\ln\left[\frac{a}{\omega c}\right]-\text{arg}\left[\Gamma\left[\frac{2i\nu c}{a}\right]\right].
\end{split}
\end{equation}
In the last line of eq.(\ref{1.44}), we use $\text{arg}\left[\Gamma[i \alpha]\right]=-\text{arg}\left[\Gamma[-i \alpha]\right]$ ($\forall ~\alpha\in\mathbb{R}$). 
The third integral in eq.(\ref{1.40}) is given by, 
\begin{equation}\label{1.45}
I'_3=-\frac{2e^{-\frac{i\omega z_0}{c}}}{\omega\left(\frac{2\omega z_0}{c}\right)^{\frac{i\nu c}{a}}}\int_0^{\frac{2\omega z_0}{c}}dx~e^{ix}x^{\frac{i\nu c}{a}}\left(1-\frac{cx}{2\omega z_0}\right)^{-\frac{i\nu c}{a}}~.
\end{equation} 
%
%
%
%
The form of the integral in eq.(\ref{1.45}) takes the following form
\begin{equation}\label{A1}
\begin{split}
I_3'=&-\frac{2e^{-\frac{i\omega z_0}{c}}}{\omega \left(\frac{2\omega z_0}{c}\right)^{\frac{i\nu c}{a}}}\int_0^{\frac{2\omega z_0}{c}}dx e^{ix}x^{\frac{i\nu c}{a}}\left(1-\frac{cx}{2\omega z_0}\right)^{-\frac{i\nu c}{a}}\\
=&-\frac{4\pi\nu z_0}{a\sinh\left[\frac{\pi \nu c}{a}\right]}e^{\frac{-i\omega z_0}{c}}\biggr[~_2\!F_3\left[a_1,a_2;\frac{1}{2},1,\frac{3}{2 };-\frac{\omega^2z_0^2}{c^2}\right]\\&+\frac{i \omega z_0}{c}\left(1+\frac{i\nu c}{a}\right) ~_2\!F_3\left[a_2,a_3;\frac{3}{2},\frac{3}{2},2;-\frac{\omega^2z_0^2}{c^2}\right]\biggr]~,
\end{split}
\end{equation}
where $a_1=\frac{1}{2}+\frac{i\nu c}{2 a}$, $a_2=1+\frac{i\nu c}{2 a}$, $a_3=\frac{3}{2}+\frac{i\nu c}{2 a}$, and $~_2\!F_3$ 
denotes the generalized hypergeometric function.
%
%
%
%
In the above equation we write, 
\begin{equation}\label{A2}
\begin{split}
\mathcal{B}_\mathcal{f}=&~_2\!F_3\left[a_1,a_2;\frac{1}{2},1,\frac{3}{2 };-\frac{\omega^2z_0^2}{c^2}\right]\\&+\frac{i \omega z_0}{c}\left(1+\frac{i\nu c}{a}\right) ~_2\!F_3\left[a_2,a_3;\frac{3}{2},\frac{3}{2},2;-\frac{\omega^2z_0^2}{c^2}\right]~,
\end{split}
\end{equation}
%
%
where $\mathcal{B}_{\mathcal{f}}$ can be expressed as, 
\begin{equation}\label{A3}
\mathcal{B}_\mathcal{f}=\left|\mathcal{B}_\mathcal{f}\right|e^{i\zeta}
\end{equation}
where $\zeta=\text{arg}\left[\mathcal{B}_\mathcal{f}\right]$. 
%
%
%
%
Adding Eqs. (\ref{A1}) and (\ref{1.43}) we obtain, 
\begin{equation}\label{A4}
\begin{split}
&I_1'+I_2'+I_3'\\&=-\frac{4\nu c}{a\omega}e^{-\frac{\pi\nu c}{a}}\left|\Gamma\left[-\frac{2i\nu c}{a}\right]\right|\cos(\theta')-\frac{4\pi \nu z_0}{a\sinh\left[\frac{\pi\nu c}{a}\right]}e^{-\frac{i\omega z_0}{c}}\mathcal{B}_{\mathcal{f}}\\
&=-\frac{4\nu c}{a\omega}\sqrt{\frac{\pi a}{\nu c}}\frac{\cos(\theta')}{\sqrt{e^{\frac{4\pi\nu c}{a}}-1}}\left[1+\frac{2\,\chi\,\omega z_0}{c}\,\frac{\mathcal{B}_\mathcal{f}\sec(\theta')e^{-\frac{i\omega z_0}{c}}}{1-e^{-\frac{2\pi\nu c}{a}}}\right]
\end{split}
\end{equation}
where $\chi=\sqrt{\frac{2\pi\nu c}{a}\sinh\left[\frac{2\pi\nu c}{a}\right]}$.
%
%
 %
 %
%
%
Using eq.(\ref{A4}) in eq.(\ref{1.42}), we can rewrite the exact analytical form of the transition probability as follows,
\begin{equation}\label{A5}
\begin{split}
\mathcal{P}_{exc}'&=\frac{16\pi\,\mathcal{G}^2\,\nu c}{a\,\omega^2}\frac{\cos^2(\theta')}{e^{\frac{4\pi\nu c}{a}}-1}\,\,\biggr[1+\frac{4\,|\mathcal{B}_\mathcal{f}|\,\chi \omega \, z_0}{c}\frac{\sec(\theta')}{1-e^{-\frac{2\pi\nu c}{a}}}\\&\times\cos\left(\zeta - \frac{\omega z_0}{c}\right)+\frac{4\,|\mathcal{B}_\mathcal{f}|^2\chi^2\omega^2z_0^2}{c^2}\frac{\sec^2(\theta')}{(1-e^{-\frac{2\pi\nu c}{a}})^2}\biggr]~.
\end{split}
\end{equation}
%
%
%
%
In order to examine the equivalence we further simplify the above equation and consider the leading order behaviour of the same. However in sec.(\ref{pheno}) we show that the exact (Eq.(\ref{A5})) and leading order (Eq.(\ref{1.47})) analysis yield same order of estimation for the transition probability.
Following the experimental proposition as considered in \cite{Flat2, Fulling}, one may take $\omega$ to be $\sim 1$ gHz. The fixed position of the atom $(z_0)$ can assumed to be a small parameter such as $ \mathcal{O}(0.01)$ m.  This leads us to write, $\frac{2 \omega z_0}{c} < 1$, where $c = 3 \times 10^8$ m/s.  
In Eq.(\ref{1.45}), the upper limit on $x$ dictates that $x < \frac{2\omega z_0}{c}$, which implies $\frac{c x}{2\omega z_0} < 1$. 
This allows us to write the Taylor expansion in the term $\left(1-\frac{cx}{2\omega z_0}\right)^{-\frac{i\nu c}{a}}$ and as well as the expansion of $e^{i x}$ upto the linear order in $x$. This yields,   
\begin{equation}\label{1.46}
\begin{split}
I'_3\cong&-\frac{2e^{-\frac{i\omega z_0}{c}}}{\omega\left(\frac{2\omega z_0}{c}\right)^{\frac{i\nu c}{a}}}\int_0^{\frac{2\omega z_0}{c}}dx~x^{\frac{i\nu c}{a}}\left(1+ix+\frac{i\nu c^2x}{2a\omega z_0}\right)\\
\cong&-\frac{4z_0}{c}e^{-\frac{i\omega z_0}{c}}\left(\frac{1}{1+\frac{i \nu c}{a}}+\frac{2i\omega z_0}{c}\frac{1+\frac{\nu c^2}{2a\omega z_0}}{2+\frac{i\nu c}{a}}\right)~.
\end{split}
\end{equation} 
Combining the Eqs.(\ref{1.43},\ref{1.46}), we get the excitation probability as follows, 
\begin{equation}\label{1.47}
\begin{split}
\mathcal{P}'_{exc}&\cong\frac{16\pi\mathcal{G}^2\nu c}{a\omega^2}\frac{\cos^2(\theta')}{e^{\frac{4\pi\nu c}{a}}-1}\biggr[\biggr(1+\frac{a\omega z_0}{\nu c^2}\frac{e^{\frac{\pi\nu c}{a}}\sec(\theta')}{\left(1+\frac{\nu^2c^2}{a^2}\right)}\\&\times\frac{\left(\cos\left(\frac{\omega z_0}{c}\right)-\frac{\nu c}{a}\sin\left(\frac{\omega z_0}{c}\right)\right)}{\left|\Gamma\left[-\frac{2i\nu c}{a}\right]\right|}\biggr)^2
\\&+\biggr(\frac{a\omega z_0}{\nu c^2}\frac{e^{\frac{\pi\nu c}{a}}\sec(\theta')}{\left(1+\frac{\nu^2c^2}{a^2}\right)}\frac{\left(\frac{\nu c}{a}\cos\left(\frac{\omega z_0}{c}\right)+\sin\left(\frac{\omega z_0}{c}\right)\right)}{\left|\Gamma\left[-\frac{2i\nu c}{a}\right]\right|}\biggr)^2\biggr]~.
\end{split}
\end{equation}
%
%
The final forms of the excitation probabilities corresponding to the two cases is summarised in Eqs.(\ref{1.29},\ref{1.47}). 
\section{Non-equivalence in dual photon scenario }\label{equivalence}
In this section we examine the equivalence between the two systems which are discussed in sections (\ref{moving_atom}, \ref{moving_mirror}), in the context of the dual photon emission. 
In our case excitation of the ground state atom (equivalent to $\hbar \omega$) is accompanied by the simultaneous emission of two photons with frequency $\nu$ (equivalent to a total $2\hbar\nu$ amount of energy). Therefore, 
following \cite{Fulling}, we interchange the atomic frequency $(\omega)$ with twice the frequency of the emitted photons $(2 \nu)$ for our setup.  
We interchange $\nu=\frac{\omega}{2}$ in Eqs.(\ref{1.29}, \ref{1.47})  respectively and obtain the two probabilities as below, 
\begin{equation}\label{1.50}
\begin{split}
\mathcal{P}_{exc}=&\frac{8\pi\mathcal{G}^2c}{a\omega}\frac{\cos^2(\bar{\theta})}{e^{\frac{2\pi\omega c}{a}}-1}\biggr(1-\frac{4\sec(\bar{\theta})}{\left|\Gamma\left(\frac{i\omega c}{a}\right)\right|}K_{\frac{i\omega c}{a}}\left(\frac{\omega c}{a}\right)\\+&\frac{4\omega c \sec^2(\bar{\theta})}{a\pi}\biggr|K_{\frac{i\omega c}{a}}\left(\frac{\omega c}{a}\right)\biggr|^2\sinh\left(\frac{\pi\omega c}{a}\right)\biggr)
\end{split}
\end{equation}
and
\begin{equation}\label{1.48}
\begin{split}
\mathcal{P}'_{exc}&\cong\frac{8\pi\mathcal{G}^2 c}{a\omega}\frac{\cos^2(\theta'')}{e^{\frac{2\pi\omega c}{a}}-1}\biggr[\biggr(1+\frac{2a z_0}{c^2}\frac{e^{\frac{\pi\omega c}{2a}}\sec(\theta'')}{\left(1+\frac{\omega^2c^2}{4a^2}\right)}\\&\times\frac{\left(\cos\left(\frac{\omega z_0}{c}\right)-\frac{\omega c}{2a}\sin\left(\frac{\omega z_0}{c}\right)\right)}{\left|\Gamma\left[-\frac{i\omega c}{a}\right]\right|}\biggr)^2
\\&+\biggr(\frac{2a z_0}{ c^2}\frac{e^{\frac{\pi\omega c}{2a}}\sec(\theta')}{\left(1+\frac{\omega^2c^2}{4a^2}\right)}\frac{\left(\frac{\omega c}{2a}\cos\left(\frac{\omega z_0}{c}\right)+\sin\left(\frac{\omega z_0}{c}\right)\right)}{\left|\Gamma\left[-\frac{i\omega c}{a}\right]\right|}\biggr)^2\biggr]~.
\end{split}
\end{equation}
Here, 
\begin{equation}\label{1.51}
\bar{\theta}=\frac{\omega z_0}{c}-\frac{\omega c}{a}\ln\left[\frac{a}{\omega c}\right]-\text{arg}\left[\Gamma\left(\frac{i\omega c}{a}\right)\right]~
\end{equation}
and 
\begin{equation}\label{1.49}
\theta''=\frac{\omega z_0}{c}-\frac{\omega c}{a}\ln\left[\frac{a}{\omega c}\right]-\text{arg}\left[\Gamma\left[\frac{i\omega c}{a}\right]\right]~.
\end{equation}
Comparing Eq.(\ref{1.48}) with Eq.(\ref{1.50}), we observe that the excitation probabilities, although has a same Planckian pre-factor,  the terms in the parenthesis do not quite match. This  indicates that the probabilities are not related to each other under such exchanges and therefore may imply a violation of the equivalence relation. This is in line with the observation made in \cite{Fulling} where a subtle symmetry between the excitation of a stationary atom due to an accelerating mirror in Minkowski spacetime with the excitation of an accelerating atom in Minkowski spacetime relative to a stationary mirror was shown. This symmetry was regarded as a manifestation of the equivalence principle. Although, there is a lot of debate about the status of the equivalence principle applying to non-gravitational processes in a gravitational field \cite{Vallisneri,Fulling0}, however, the observation made in \cite{Fulling} indeed shows a nice symmetry between the two set ups in the one photon case which is found to be broken in our case. 
%
%
Also from the exact result {\it i.e} Eq.(\ref{A5}) one can obtain the same insight leading to the violation of equivalence relation.
%
%
%
%
In the upcoming section we study the observational prominence of the dual photon emission scenario. 
\section{Phenomenological aspects of the model}\label{pheno}
\begin{figure}[h!]
    \includegraphics[scale=0.31]{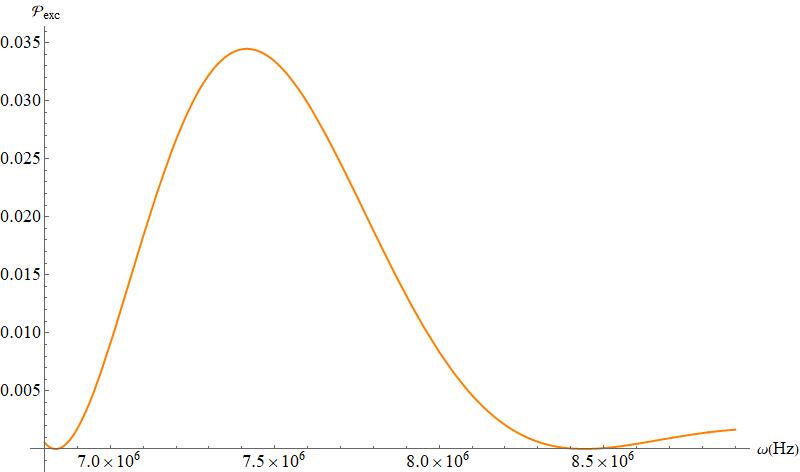}
    \caption{Plot of $\mathcal{P}_{exc}$ vs. $\omega$. Atom is accelerating with respect to the fixed mirror and two photons are simultaneously emitted. }
    \label{atom_accl}
\end{figure}

In this section we study the behaviour of the system while the atom is excited to its higher energy state along with the simultaneous emission of two photons. Therefore in fig.(\ref{atom_accl}), we plot the transition probability 
as depicted in Eq.(\ref{1.29}) 
with respect to the angular frequency of the detector $(\omega)$. We fix the parameters such as, $\nu \sim 10^{4}$ Hz, $z_0= 0.01$ m. We also take the effective coupling strength between the $N$ $(\sim 10^4-10^8$) polar molecules and the photons as, $g_{{\rm eff}}= g \sqrt{N} \sim 10^7$ Hz. The acceleration of the atom is fixed to be $\sim 10^{15}$ m/s$^2$, which can be achieved by using the superconducting circuits \cite{Friis2012cx}.
Fig.(\ref{atom_accl}) depict the standard Planckian distribution and also shows an oscillatory behaviour as similar to the single photon emission case.

\begin{figure}[h!]
    \includegraphics[scale=0.355]{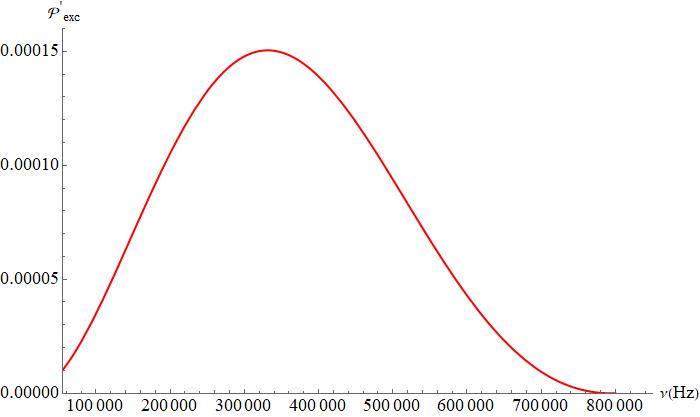}
    \caption{Plot of $\mathcal{P}'_{exc}$ vs. $\nu$. Mirror is accelerating with respect to a fixed atom and two photons are simultaneously emitted. }
    \label{mirror_accl_1}
\end{figure}

\begin{figure}[h!]
\includegraphics[scale=0.35]{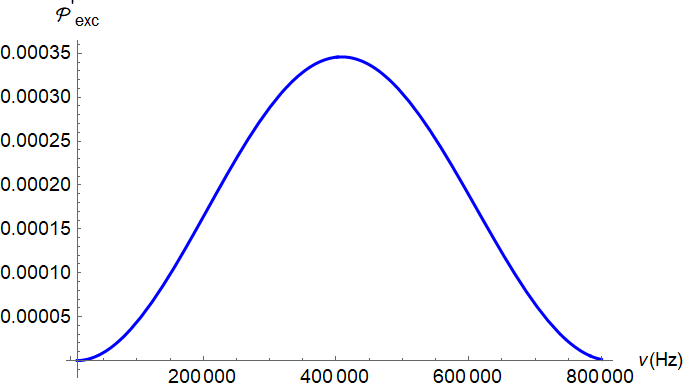} 
\caption{Plot of $\mathcal{P}'_{exc}$ vs. $\nu$. for $\frac{\nu c}{a}\ll 1$}
\label{mirror_accl}
%
 \end{figure}

Now we 
explore the other system which is the mirror accelerating with respect to a fixed atom. 
%
%
%
Largely, we fix the values for the parameters as considered in \cite{Fulling}, so that 
the observational prominence of our model may also be verified with respect to the similar experimental setup as proposed in \cite{Fulling}.  
 This leads us to fix $\omega \sim 10^{9}$ Hz, $z_0= 0.01$ m,  $g_{{\rm eff}}= g \sqrt{N} \sim 10^7$ Hz and acceleration of the mirror to be $\sim 10^{15}$ m/s$^2$ in eq.(\ref{A5}) and obtain the behaviour of the transition probability with the variation of the field frequency $(\nu)$ in Fig. (\ref{mirror_accl_1}). 
 %
 %
This plot suggests that in case of the dual photon emission process, the transition probability becomes $\mathcal{O}(10^{-4})$ which is similar to the outcome of single field emission. We comment that in case of the dual photon emission the excitation probability can be enhanced to the detectable probability ({\it i.e} $P \sim \mathcal{O}(10^{-2})$) \cite{Fulling} by taking into account the probabilities for $100$ such cavity modes. We can also observe similar behaviour of the approximate form of the excitation probability in eq.(\ref{1.47}) if we use the same parameters as in Fig.(\ref{mirror_accl_1}).
Now we analyse Eq.(\ref{1.48}) under the condition $\frac{c\nu}{a}\ll1$. We keep fix rest of the parameters  $(\omega,z_0,g_{{\rm eff}},a)$ same as taken in the Fig.(\ref{mirror_accl_1}). This yields, 
\begin{equation}\label{approx_1}
\begin{split}
&\mathcal{P}'_{exc}|_{\nu c \ll a}\,\cong\frac{4\mathcal{G}^2\,\cos^2(\theta')}{\omega^2}\biggr[1+\frac{2 a \omega z_0}{\nu c^2}\frac{\sec(\theta')}{\left|\Gamma\left[-\frac{2i\nu c}{a}\right]\right|}\\
&+\biggr(\frac{a \omega z_0}{\nu c^2}\frac{\sec(\theta')}{\left|\Gamma\left[-\frac{2i\nu c}{a}\right]\right|}\biggl)^2\biggr\{1+\frac{2\pi\nu c}{a}\biggl(1+\frac{1}{\pi}\biggr) \sin\biggr(\frac{\omega z_0}{c}\biggl)\\&+\sin^2\biggr(\frac{\omega z_0}{c}\biggl)\biggl\}\biggl]~.
\end{split}
\end{equation}
We further plotted Eq.(\ref{approx_1}) with respect to $\nu$ and obtained the Fig.(\ref{mirror_accl}). 
%
%
%
\section{Discussion}\label{discussion}
Transforming a virtual photon to a real detectable photon has become a remarkable tool in studying the acceleration radiation and its several consequences in the flat/curved spacetime. Besides its theoretical aspects, this mechanism is also promising from the observational perspective. 
In the present manuscript we consider an atom-mirror system where the atom and the mirror possess relative acceleration between them.  Two configurations are possible such as the atom is accelerating with respect to a fixed mirror and vice versa. 
We further consider that due to the virtual processes, which take place in the vacuum states of the two scalar photons, the atom makes a transition to its higher excited state along with the simultaneous emission of the two virtual photons. Applying the mechanism of transforming the virtual photons to real one, we evaluate the transition probabilities corresponding to the above configurations. 
Our results turn out to be distinct in nature than that of the single field emission case. 
We obtain that the dual field emission process breaks the ``equivalence principle in QED".  An interference term emerges in the transition probability due to the emission of the two photon quanta, which in turn breaks the equivalence between the two setup. Undoubtedly, this demands more such investigations to perceive the notion, applications and the limitations of such equivalence principle.  
As similar to the single filed emission process, our results suggest that both the transition probabilities as in Eqs.(\ref{1.29}, \ref{1.47}) exhibit spatial oscillatory behaviour.  An additional interference term excluding the standard Planckian factor is solely responsible for this oscillatory behaviour in the excitation probability.
For our case too, a similar experimental proposition as mentioned in \cite{Fulling} can be constructed as following. 
One may consider that $N (\sim 10^4 - 10^8)$ number of polar molecules are collectively forming an ensemble where each molecule is simultaneously interacting with the two photons within a cavity setup. We refer our readers to \cite{Fulling,wilson_nature,Flat1,Flat2}
for a thorough understanding of the experimental setup which demonstrate the motion of accelerated mirror and capture its several consequences. 
Note that for mathematical simplicity we translate the moving mirror and static atom case in Minkowski spacetime to  Rindler spacetime, where  the mirror becomes static and the atom accelerates. This permits us to  define a Rindler vacuum state with respect to the static mirror in Rindler spacetime. 
In this context, we point out that it is a very difficult or somewhat impossible task to construct a Rindler vacuum state in the laboratory. For this reason, the experimental implementation of this setup (which is briefly mentioned earlier) is constructed in the Minkowski spacetime where the mirror accelerates and the atom is static \cite{Fulling,wilson_nature,Flat1,Flat2}. 

We observe that for an accelerating mirror system, under the condition $a \gg c \nu$, the transition probability of the system continues to be the function of $(a,\, c, \nu)$. This implies that the parameter space in the dual photon emission process is larger than that of the single field emission case, even under the above condition. Therefore we gain the freedom to study the transition probability with respect to the field frequency, acceleration of the mirror etc. 
%
%
%
Further, we obtain a detectable transition probability by conveniently choosing the range of field frequency and the acceleration of the mirror with respect to the fixed atom. 
%
%
%
%
 It will be interesting to extended the present work with the emission of more than two rather larger number of photons. We are planning to report this elsewhere. 


\begin{thebibliography}{8}
\bibitem{Einstein15}
A. Einstein, Sitzungsber Preuss Akad Wiss (1915) 844.
\bibitem{Einstein16}
A. Einstein, \href{https://doi.org/10.1002/andp.19163540702}{Ann. der Physik 49 (1916) 769}.
\bibitem{Hawking}
S.W. Hawking, \href{https://doi.org/10.1038/248030a0}{Nature 248 (1974) 30}.
\bibitem{Hawking2}
S.W. Hawking, \href{https://doi.org/10.1007/BF02345020}{Commun. Math. Phys 43 (1975) 199}.
\bibitem{Hawking3}
S.W. Hawking, \href{https://link.aps.org/doi/10.1103/PhysRevD.13.191}{Phys. Rev. D 13 (1976) 191}.
\bibitem{Bekenstein}
J.D. Bekenstein, \href{https://doi.org/10.1007/BF02757029}{Lett. Nuovo Cimento 4 (1972) 737}.
\bibitem{Bekenstein2}
J. D. Bekenstein, \href{https://link.aps.org/doi/10.1103/PhysRevD.7.2333}{Phys. Rev. D 7 (1973) 2333}.
\bibitem{Page}
D. N. Page, \href{https://link.aps.org/doi/10.1103/PhysRevD.13.198}{Phys. Rev. D 13 (1976) 198}.
\bibitem{Page2}
D. N. Page, \href{https://link.aps.org/doi/10.1103/PhysRevD.14.3260}{Phys. Rev. D 14 (1976) 3260}.
\bibitem{Page3}
D. N. Page, \href{https://link.aps.org/doi/10.1103/PhysRevD.16.2402}{Phys. Rev. D 16 (1977) 2402}.
\bibitem{Fulling21}
S. A. Fulling, \href{https://link.aps.org/doi/10.1103/PhysRevD.7.2850}{Phys. Rev. D 7 (1973) 2850}.
\bibitem{Davies}
P. Davies, \href{https://iopscience.iop.org/article/10.1088/0305-4470/8/4/022}{J. Phys. A 8 (1975) 609}.
\bibitem{DeWitt}
B. S. DeWitt, General Relativity: An Einstein Centenary Survey, eds S. W. Hawking,
W. Israel (Cambridge Univ Press (1979), Cambridge, UK).
%
%
\bibitem{Scully2003zz}
M. O. Scully, V. V. Kocharovsky, A. Belyanin, E. Fry and F. Capasso,\href{https://link.aps.org/doi/10.1103/PhysRevLett.91.243004}{Phys. Rev. Lett. 91 (2003) 243004}.
%
%
\bibitem{Fulling2}
M. O. Scully, S. A. Fulling, D. M. Lee, D. N. Page, W. P. Schleich and A. A. Svidzinsky, \href{https://doi.org/10.1073/pnas.1807703115}{Proc. Natl. Acad. Sci. U.S.A. 115 (2018) 8131}.
\bibitem{Fulling}
A. A. Svidzinsky, S. J. Ben-Benjamin, S. A. Fulling and D. N. Page, \href{https://link.aps.org/doi/10.1103/PhysRevLett.121.071301}{Phys. Rev. Lett 121 (2018) 071301}.
\bibitem{Ordonez1}
H. E. Camblong, A. Chakraborty and C. R. Ord\'o\~nez, \href{https://link.aps.org/doi/10.1103/PhysRevD.102.085010}{Phys. Rev. D 102 (2020) 085010}.
\bibitem{Ordonez2}
A. Azizi, H. E. Camblong, A. Chakraborty, C. R. Ord\'o\~nez and M. O. Scully, \href{https://link.aps.org/doi/10.1103/PhysRevD.104.065006}{Phys. Rev. D 104 (2021) 065006}.
%
%
\bibitem{Ordonez3}
A. Azizi, H. E. Camblong, A. Chakraborty, C. R. Ord\'o\~nez and M. O. Scully, \href{https://link.aps.org/doi/10.1103/PhysRevD.104.084086}{Phys. Rev. D 104 (2021) 084086}.
%
%
\bibitem{Ordonez4}
A. Azizi, H. E. Camblong, A. Chakraborty, C. R. Ord\'o\~nez and M. O. Scully, \href{https://link.aps.org/doi/10.1103/PhysRevD.104.084085}{Phys. Rev. D 104 (2021) 084085}.
%
%
\bibitem{OTM}
S. Sen, R. Mandal and S. Gangopadhyay, \href{https://link.aps.org/doi/10.1103/PhysRevD.105.085007}{Phys. Rev. D 105 (2022) 085007}.
%
%
\bibitem{OTM2}
S. Sen, R. Mandal and S. Gangopadhyay, \href{https://link.aps.org/doi/10.1103/PhysRevD.106.025004}{Phys. Rev. D 106 (2022) 025004}.
%
%
\bibitem{GWGibbons}
G. W. Gibbons and S. W. Hawking, \href{https://link.aps.org/doi/10.1103/PhysRevD.15.2752}{Phys. Rev. D 15 (1977) 2752}.
%
%
\bibitem{Schwinger}
J. Schwinger, \href{https://link.aps.org/doi/10.1103/PhysRev.82.664}{Phys. Rev. 82 (1951) 664}.
%
%
\bibitem{Nielsen}
M. A. Nielsen and I. A. Chuang, ``\textit{Quantum Computation and Quantum Information: 10th Anniversary Edition}" (2011), Cambridge University Press, United States.
%
%
\bibitem{Unruh21}
W. G. Unruh, \href{https://link.aps.org/doi/10.1103/PhysRevD.14.870}{Phys. Rev. D 14 (1976) 870}.
%
%
\bibitem{Unruh22}
W. G. Unruh, \href{https://link.aps.org/doi/10.1103/PhysRevD.15.365}{Phys. Rev. D 15 (1977) 365}.
%
%
\bibitem{Unruh2}
W. G. Unruh and R. M. Wald, \href{https://link.aps.org/doi/10.1103/PhysRevD.29.1047}{Phys. Rev. D 29 (1984) 1047}.
%
%
\bibitem{Higuichi}
L. C. B. Crispino, A. Higuchi, G. E. A. Matsas, \href{https://link.aps.org/doi/10.1103/RevModPhys.80.787}{Rev. Mod. Phys. 80 (2008) 787}.
%
%
\bibitem{Cosmospace}
Aindri\'{u} Conroy,\href{https://link.aps.org/doi/10.1103/PhysRevD.105.123513}{Phys. Rev. D 105 (2022) 123513}.
%
%
 \bibitem{Birrell} 
  N.~D.~Birrell and P.~C.~W.~Davies,
 ``Quantum Fields in Curved Space,'' (Cambridge University Press, Cambridge) 1982.
%
%
\bibitem{SG}
R. Chatterjee, S. Gangopadhyay and A. S. Majumdar, \href{https://link.aps.org/doi/10.1103/PhysRevD.104.124001}{Phys. Rev. D 104 (2021) 124001}.
%
%
\bibitem{wilson_nature}
C. M. Wilson, G. Johansson, A. Pourkabirian, M. Simoen, J. R. Johansson, T. Duty, F. Nori, and P. Delsing, \href{https://www.nature.com/articles/nature10561}{Nature (London) 479, (2011) 376}.

\bibitem{Flat1}
J. R. Johansson, G. Johansson, C. M. Wilson and F. Nori, \href{https://link.aps.org/doi/10.1103/PhysRevLett.103.147003}{Phys. Rev. Lett. 103 (2009) 147003}.
%
%
\bibitem{Flat2}
M. Wallquist, K. Hammerer, P. Rabl, M. Lukin and P. Zoller, \href{http://dx.doi.org/10.1088/0031-8949/2009/T137/014001}{Phys. Scr. T137 (2009) 014001}.
%
%
\bibitem{Das}
C. Chowdhury, A. Das and B. R. Majhi, \href{https://doi.org/10.1140/epjp/i2019-12400-2}{Eur. Phys. J. Plus 134 (2019) 65}.
%
%
\bibitem{Davoudiasl:1999jd}
H.~Davoudiasl, J.~L.~Hewett and T.~G.~Rizzo,
Phys. Rev. Lett. \textbf{84}, 2080 (2000)
%
%

\bibitem{Chang:1999nh}
S.~Chang, J.~Hisano, H.~Nakano, N.~Okada and M.~Yamaguchi,
Phys. Rev. D \textbf{62}, 084025 (2000)

\bibitem{Dienes:2002hg}
K.~R.~Dienes,
2002 TASI Lectures, ``New directions for new dimensions: An introduction to Kaluza-Klein theory, large extra dimensions, and the brane world''
%
%

\bibitem{Dvali:2007hz} 
  G.~Dvali, Fortsch.\ Phys.\  {\bf 58}, 528 (2010)
  
  \bibitem{Dvali:2009ne} 
  G.~Dvali and M.~Redi, Phys.\ Rev.\ D {\bf 80}, 055001 (2009)
%
%

 \bibitem{Dvali:2009fw} 
  G.~Dvali, I.~Sawicki and A.~Vikman, JCAP {\bf 0908}, 009 (2009)
  %
  %
  
\bibitem{Wands:2007bd}
D.~Wands,
Lect. Notes Phys. \textbf{738}, 275-304 (2008)
%
%

\bibitem{Gong:2016qmq}
J.~O.~Gong,
Int. J. Mod. Phys. D \textbf{26}, no.01, 1740003 (2016)

\bibitem{Alvarez-Gaume:2011wmd}
L.~Alvarez-Gaume, C.~Gomez and R.~Jimenez,
JCAP \textbf{03}, 017 (2012)
 %
 %
 
\bibitem{NooriGashti:2021nox}
S.~Noori Gashti,
JHAP \textbf{2}, no.1, 13-24 (2022).
%
%

\bibitem{Martin:2021frd}
J.~Martin and L.~Pinol,
JCAP \textbf{12}, no.12, 022 (2021).
%
%
\bibitem{Akrami:2020zfz}
Y.~Akrami, M.~Sasaki, A.~R.~Solomon and V.~Vardanyan,
Phys. Lett. B \textbf{819}, 136427 (2021)
%
%
\bibitem{Philbin}
T. G. Philbin, et al., \href{http://dx.doi.org/10.1126/science.1153625}{Science 319 (2008) 1367}.
%
%
\bibitem{Rohrlich}
F. Rohrlich, \href{https://doi.org/10.1016/0003-4916(63)90051-4}{Ann. Phys. 22 (1963) 169}.
\bibitem{Vallisneri}
M. Pauri and M. Vallisneri, \href{https://doi.org/10.1023/A:1018821619763}{Foun. Phys. 29 (1999) 1499}.
\bibitem{Singleton}
D. Singleton and S. Wilburn, \href{https://link.aps.org/doi/10.1103/PhysRevLett.107.081102}{Phys. Rev. Lett. 107 (2011) 081102}.
\bibitem{Singleton2}
D. Singleton and S. Wilburn, \href{https://doi.org/10.1142/S0218271816440090}{Int. J. Mod. Phys. D 25 (2016) 1644009}.
\bibitem{Fulling0}
S. A. Fulling and J. H. Wilson, \href{https://iopscience.iop.org/article/10.1088/1402-4896/aaecaa}{Phys. Scr. 94 (2019) 014004}.
%
%
\bibitem{Gradshteyn}
I. S. Gradshteyn and I. M. Ryzhik, \textit{Table of Integrals, Series and Products}, 7th edition (Academic Press, Amsterdam, 2007).
\bibitem{HAMLD}
F. Hammad, A. Landry and D. Dijamco, \href{https://link.aps.org/doi/10.1103/PhysRevD.103.085010}{Phys. Rev. D 102 (2021) 085010}.
\bibitem{Sue}
D. Su, C. T. M. Ho, R. B. Mann, and T. C. Ralph, \href{https://iopscience.iop.org/article/10.1088/1367-2630/aa71d1}{New J. Phys. 19 (2017) 063017}.
\bibitem{Friis2012cx}
N. Friis, A. R. Lee, K. Truong, C. Sabin, E. Solano, G. Johansson and I. Fuentes,\href{https://link.aps.org/doi/10.1103/PhysRevLett.110.113602}{Phys. Rev. Lett. 110 (2013) 113602}.
\end{thebibliography}
\end{document}